\documentclass[aps,prd,twocolumn,showpacs,showkeys,superscriptaddress]{revtex4-2}
\usepackage{latexsym}
\usepackage{amssymb}
\usepackage{amsmath}
\usepackage{amscd}
\usepackage{amsthm}
\usepackage{graphicx}
\usepackage{textcomp}
\usepackage{colortbl}
\usepackage[colorlinks]{hyperref}
\usepackage[font={footnotesize,it}]{caption}
\usepackage{multirow}
\usepackage[T1]{fontenc}
\usepackage{ae,aecompl}
\usepackage{subcaption}
\usepackage{orcidlink}
\begin{document}
	
	\title{Generalized Second Law and Thermodynamical Aspects of $f(Q,\mathcal{T})$ Gravity}
	\vspace{10mm}
		
	\author{Anirudh Pradhan \orcidlink{0000-0002-1932-8431}
	}
	\email{pradhan.anirudh@gmail.com}
	\affiliation{Centre for Cosmology, Astrophysics and Space Science (CCASS),
		GLA University, Mathura-281406, U.P., India.}
	
	\author{A. Husain \orcidlink{0000-0002-0231-9395}}
	\email[]{atherhusain1001@gmail.com}
	\affiliation{Department of Mathematics, Narayanrao kale Smurti Model College Karanja (Gha), Wardha, Maharashtra 442203, India}
	\author{M. Zeyauddin\orcidlink{0000-0001-8382-8994}}
	\email[]{uddin\_m@rcjy.edu.sa}
	\affiliation{Department of General Studies (Mathematics) Jubail Industrial College, Jubail 31961, Saudi Arabia}
	\author{S. H. Shekh \orcidlink{0000-0003-4545-1975}}
	\email[]{da\_salim@rediff.com}
	\affiliation{Department of Mathematics, S.P.M. Science and Gilani Arts, Commerce College, Ghatanji, Yavatmal, \\Maharashtra-445301, India}
	\affiliation{Pacif Institute of Cosmology and Selfology (PICS), Sagara, Sambalpur 768224, Odisha, India}
	\vspace{10mm}
	\begin{abstract}
		\textbf{Abstract} Late-time cosmic acceleration has motivated the exploration of various extensions of general relativity, among which $f(Q,\mathcal{T})$ gravity, based on the non-metricity scalar $Q$ and the trace of the energy--momentum tensor $\mathcal{T}$, has gained increasing attention. In this study, we explore the thermodynamic aspects of  $f(Q,\mathcal{T})$ gravity by establishing the first law and generalized second law of thermodynamics at the apparent horizon of a flat FLRW universe. By applying the Gibbs relation, we determined the rate of change of the total entropy and assessed the conditions under which the generalized second law remains valid for various choices of  $f(Q,\mathcal{T})$. Our analysis focuses on linear, power-law, quadratic trace, exponential, and cross-coupling models, inspired by frameworks such as $f(R,\mathcal{T})$, $f(T)$, and modifications motivated by string theory.  Our analysis shows a clear trend that the linear and weakly nonlinear models remain thermodynamically consistent. Strongly nonlinear and interaction-based models both are more sensitive and in need of careful tuning of parameters,  otherwise, the generalized second law may fail. This concepts is an important point. Thermodynamics can act as a useful test. \\
		\noindent
\textbf{keywords:} {Modified $f(Q,T)$ gravity; Thermodynamics; Generalized second law; Cosmic acceleration; Entropy evolution}		
	\end{abstract}
	\maketitle
	\section{Introduction}
	
The late-time acceleration of the universe is a key result in modern cosmology. It is supported by many observations. These include Type Ia supernovae, cosmic microwave background data, and baryon acoustic oscillations. In the standard model, this acceleration is explained by dark energy. This form of energy makes up about 70\% of the total cosmic content. Many models have been proposed to describe it. These include the cosmological constant, quintessence, k-essence, Chaplygin gas, and holographic dark energy \cite{Copeland2006,Padmanabhan2003,Heisenberg2019}. However, these models face several problems. They often require fine tuning. A clear theoretical foundation is still missing. Some also struggle to match all observations. Because of this, researchers look for other ideas. One important approach is to modify the theory of gravity itself.

Modified gravity models offer an alternative to dark energy, which accounts for cosmic acceleration through modifications of the spacetime geometry. The most basic and extensively analyzed formulation in this class is $f(R)$ gravity, defined by replacing the Ricci scalar in the Einstein–Hilbert action with the general function $f(R)$ \cite{Sotiriou2010,DeFelice2010,A1,A2}. Another important extension is $f(T)$ gravity, which is constructed in the teleparallel framework, where torsion replaces the curvature as the fundamental entity. In particular, $f(T)$ gravity, formulated in a teleparallel framework in which torsion replaces curvature, has attracted considerable attention. Several significant studies on both the astrophysical and cosmological aspects of $f(T)$ gravity can be found in the literature \cite{Capozziello2011,Myrzakulov2011,Jeon2011,Tamanini2012,Cai2016,Anagnostopoulos2019,Nair2022,Shekh2020,Chirde2019,Chirde2018, A3,A4,A5,A6}, where issues such as cosmic acceleration, large-scale structures, black hole solutions, and observational viability have been extensively investigated. These findings underscore the richness of the $f(T)$ framework and its viability as a strong alternative to general relativity in describing late-time cosmic acceleration, with further developments such as $f(R,\mathcal{T})$ gravity proposed by Harko et al. \cite{Harko2011}, where the gravitational action depends not only on the Ricci scalar $R$ but also on the trace of the energy–momentum tensor $\mathcal{T}$. This framework naturally incorporates an explicit matter–geometry coupling, which leads to non–conservation of the energy–momentum tensor and allows for richer cosmological dynamics compared to pure $f(R)$ theories. A considerable number of studies have been devoted to exploring the astrophysical and cosmological consequences of $f(R,\mathcal{T})$ gravity \cite{A8,A9,A10,A11}. These studies cover diverse directions, including the formulation of viable dark energy models capable of explaining late-time cosmic acceleration, investigation of anisotropic cosmologies and their dynamical behavior, construction and stability analysis of wormhole solutions, and detailed examinations of energy conditions to ensure theoretical and physical viability.

In recent years, the formulation of symmetric teleparallel gravity, constructed on the non-metricity scalar $Q$, has led to the development of new frameworks, most notably $f(Q)$ gravity and its various extensions \cite{Jimenez2018,A12,A13,A14,A15,A15a}. Among these, the $f(Q,\Sigma,\mathcal{T})$ and  $f(Q,\mathcal{T})$ theories (where $\Sigma$ is a contraction of non-metricity) have been explored in many recent studies, such as in bouncing cosmologies \cite{GulSharif2024}, wormholes \cite{WormholeQTR2024}, and Weyl-type matter bounce models in  $f(Q,\mathcal{T})$ gravity \cite{Zhadyranova2024}. The role of the boundary term in $f(Q,B)$ symmetric teleparallel gravity \cite{Capozziello}. These frameworks offer novel geometric perspectives on gravitation and, in many cases, retain desirable mathematical features such as second-order field equations. This  properties are not only simplify the theoretical formulation but it enhance  physically, which act as promising candidates for explaining the cosmic acceleration without introducing unknown dark-energy components.
	
Among these extensions,  $f(Q,\mathcal{T})$ gravity \cite{A10, A16,A17,A18} has emerged as a particularly interesting model owing to the explicit coupling between the non-metricity scalar $Q$ and the trace of the energy-momentum tensor $\mathcal{T}$. This coupling introduces novel matter-geometry interactions, which can lead to effective dark energy behavior without the need for additional exotic fields. This opens the door to dynamical cosmological scenarios, providing a rich phenomenology that can account for both early- and late-time cosmic evolution. Recent investigations, such as the bouncing cosmological models in  $f(Q,\mathcal{T})$ gravity \cite{GulSharif2024} and the Weyl-type  $f(Q,\mathcal{T})$ matter bounce scenario \cite{Zhadyranova2024}, have shown that singularity avoidance, stability, and viability (in terms of energy conditions or observational behavior) can often be achieved. In addition, the construction of wormhole geometries in  $f(Q,\mathcal{T})$ with viscous matter \cite{WormholeQTR2024}, indicates that such theories can potentially satisfy the energy condition constraints that are usually violated in standard GR wormhole solutions. In view of these aspects, examining the thermodynamic consistency of the theory in greater detail becomes essential for validating its applicability.

A substantial body of work has established that gravitational dynamics in modified theories of gravity admit a consistent thermodynamic interpretation at the apparent horizon of a Friedmann–Lemaître–Robertson–Walker (FLRW) universe. In particular, it has been shown within the framework of $f(R)$ gravity that the modified field equations can be recast into a form analogous to the unified first law of thermodynamics \cite{Bamba2010},
\begin{equation}
		dE = T_h dS_h + W dV + T_h d\bar{S},
\end{equation}
where the entropy associated with the horizon is generalized as $S_h = \frac{A f_R}{4G}$, with $f_R = \partial f/\partial R$. The emergence of the additional entropy production term $d\bar{S}$ reflects the intrinsic non-equilibrium nature of thermodynamics in curvature-based extensions of general relativity. Further developments have demonstrated that an equilibrium description can also be constructed under suitable redefinitions of thermodynamic variables, allowing the generalized second law of thermodynamics (GSLT) to be satisfied across different cosmological phases, including both quintessence and phantom regimes, provided that the temperature of the cosmic fluid is consistent with that of the apparent horizon \cite{AkbarR2007}.\\	
In parallel, investigations in torsion-based $f(T)$ gravity have revealed that thermodynamic consistency can be preserved even in the presence of nontrivial cosmological phenomena such as finite-time future singularities and rip scenarios. In such models, the horizon entropy is modified as $S_h = \frac{A f_T}{4G}$, where $f_T = \partial f/\partial T$, and the validity of the GSLT is closely linked to the interplay between the torsional corrections and the thermal equilibrium condition. These studies point to a common idea. The thermodynamic behavior of modified gravity depends strongly on the geometry of spacetime. Small changes in the framework can affect the results. The generalized second law plays an important role. It acts as a strong test for these models. It helps decide whether a model is physically acceptable. \cite{BambaR2012}.

The profound connection between gravitation and thermodynamics has long been acknowledged as a fundamental guiding principle in theoretical physics, offering deep insights into the nature of spacetime, horizon dynamics, and  underlying laws governing cosmic evolution. Building on the work of Jacobson \cite{Jacobson1995}, who derived Einstein’s field equations from the Clausius relation, a profound link between gravitational dynamics and horizon thermodynamics was established. This pivotal insight has driven extensive investigations into cosmological thermodynamics across various modified gravity frameworks, including $f(R)$, $f(T)$, and $f(R,\mathcal{T})$ theories \cite{Akbar2007,Cai2005,Bamba2011,Capozziello1,Filho}. When evaluating the physical consistency of these modified theories, the key thermodynamic criteria include the generalized second law (GSLT), entropy-area relationships, and horizon temperature. However, in the context of  $f(Q,\mathcal{T})$ gravity,  thermodynamic investigations are still relatively sparse, compared with those in pure $f(Q)$ or $f(R,\mathcal{T})$ settings. Given that recent studies on  $f(Q,\mathcal{T})$ (for example, the bouncing models and wormholes cited above) emphasize the behavior of energy density, pressure, stability etc, thermodynamic laws (first and second, equilibrium vs non-equilibrium) can provide additional constraints on the functional forms of  $f(Q,\mathcal{T})$. Hence, the study of thermodynamics in  $f(Q,\mathcal{T})$ gravity is both timely and essential for probing the nature of the dark energy and the dynamics of the universe. 
	
The literature on modified gravity theories features a well-established and robust line of inquiry based on thermodynamic principles.  For instance, the foundational connection between horizon thermodynamics and gravitational field equations 	was first developed in the context of general relativity and was extended to $f(R)$ gravity 	by Eling et al.~\cite{Eling2006}. Recently, thermodynamic analyses have also been performed in the framework of 	$f(Q)$ and $f(Q,\mathcal{T})$ gravity, highlighting the role of non-metricity in driving 	late-time acceleration \cite{Xu2019}. Furthermore, reviews by Bamba and Odintsov~\cite{Bamba2012} and Nojiri et al.~\cite{Nojiri2017} emphasized that the generalized second law can act as a universal consistency criterion 	for a wide class of dark-energy and modified-gravity models. Hence, our research to this well-established thermodynamic paradigm, it is not only underscore the validity of $f(Q,\mathcal{T})$ gravity but  it is a powerful framework capable for the theoretical predictions with stringent observational constraints.
	
\section{Basic Formalism in $f(Q,\mathcal{T})$ Gravity}
In the past few years, modified gravity has developed into a broad and active area of research in cosmology. Its main goal is to address some of the fundamental challenges of the standard model of the universe, especially the observed late-time cosmic acceleration and the unclear nature of dark energy. In curvature-based approaches such as $f(R)$ gravity, the accelerated expansion is explained by modifying the geometric part of the gravitational action. On the other hand, torsion-based theories like $f(T)$ gravity provide an alternative description within the teleparallel framework. More recently, another direction has gained attention, namely symmetric teleparallel gravity, where gravitational effects arise from non-metricity instead of curvature or torsion. This approach has led to the development of models such as $f(Q)$ and its generalized forms. In addition to these developments, modern extensions often incorporate interactions between matter and geometry, which enrich the underlying dynamics and offer greater flexibility in modeling cosmic evolution. Within this context, $f(Q,\mathcal{T})$ gravity emerges as a natural and significant generalization, as it combines non-metricity with matter–geometry coupling. Although several works have examined its cosmological implications and dynamical behavior, its thermodynamic aspects have not been explored in sufficient detail. In particular, issues related to entropy evolution and the validity of the generalized second law of thermodynamics are still not fully understood. Therefore, a systematic thermodynamic investigation becomes essential, both to assess the physical consistency of the model and to establish meaningful connections with other modified gravity frameworks  \cite{NojiriOdintsov,Bahamonde,Kaczmarek}.
To construct the theoretical framework of $f(Q,\mathcal{T})$ gravity, one begins with a suitable action that generalizes symmetric teleparallel gravity by including non-minimal coupling between the non-metricity scalar $Q$ and the trace of the matter energy–momentum tensor $\mathcal{T}$. This approach naturally extends the symmetric teleparallel equivalent of general relativity (STEGR), where $f(Q,\mathcal{T}) = Q$, by allowing richer interactions between the geometry and matter sectors. Such couplings offer a mechanism to reproduce dark energy behavior, generate new dynamical phenomena, and expand the available parameter space, thereby enriching the exploration of diverse cosmological and astrophysical scenarios.
	The action for $f(Q,\mathcal{T})$ gravity is \cite{Xu2019}:
	\begin{equation}\label{1}
		S = \int \left[ \frac{1}{16\pi} f(Q,\mathcal{T}) + \mathcal{L}_m \right] \sqrt{-g} \, d^4x,
	\end{equation}
	where $f(Q,\mathcal{T})$ is an arbitrary function of $Q$ and $\mathcal{T}$, and $\mathcal{L}_m$ is the Lagrangian matter.
	
	The non-metricity scalar is
	\begin{equation}\label{2}
		Q = - g^{\mu\nu}\left(L^{\alpha}_{\phantom{\alpha}\mu\beta} L^{\beta}_{\phantom{\beta}\nu\alpha} - L^{\alpha}_{\phantom{\alpha}\mu\nu} L^{\beta}_{\phantom{\beta}\alpha\beta}\right),
	\end{equation}
	where $L^{\alpha}_{\phantom{\alpha}\mu\nu}$ is the deformation tensor.
	
	The energy--momentum tensor is defined as
	\begin{equation}\label{3}
		\mathcal{T}_{\mu\nu} = - \frac{2}{\sqrt{-g}} \frac{\delta(\sqrt{-g}\mathcal{L}_m)}{\delta g^{\mu\nu}},
	\end{equation}
	and its variation contributes to $\Theta_{\mu\nu} = g^{\alpha\beta} \delta \mathcal{T}_{\alpha\beta} / \delta g^{\mu\nu}$.
	
	The variation in the action yields the  following field equations \cite{Xu2019}:
	\begin{widetext}
		\begin{equation}\label{4}
			-2 \nabla_\alpha(f_Q P^{\alpha}_{\phantom{\alpha}\mu\nu}) - \frac{1}{2} f g_{\mu\nu} + f_{\mathcal{T}} (\mathcal{T}_{\mu\nu} + \Theta_{\mu\nu})- f_Q\left(P_{\mu\alpha\beta}Q_{\nu}^{\phantom{\nu}\alpha\beta} - 2Q^{\alpha\beta}_{\phantom{\alpha\beta}\mu}P_{\alpha\beta\nu}\right) = 8\pi \mathcal{T}_{\mu\nu}
		\end{equation}
	\end{widetext}
	where $f_Q = \partial f / \partial Q$ and $f_{\mathcal{T}} = \partial f / \partial \mathcal{T}$. The above equation  represents the modified field equations for  $f(Q,\mathcal{T})$ gravity. They clearly differ from Einstein’s equations in the presence of  derivatives $f_Q$ and $f_\mathcal{T}$, which encode the response of the action to variations in the non-metricity and matter sectors, respectively. As a result, the dynamics of spacetime are not only governed by the geometry but are also directly influenced by the trace of the energy–momentum tensor, thereby realizing a genuine matter–geometry interaction.
	
	\section{FLRW Universe and Field Equations}
To explore the cosmological consequences, we specialize the field equations in a homogeneous and isotropic universe, described by the spatially flat FLRW metric. This symmetry reduction yields a set of generalized Friedmann equations that govern the evolution of the cosmic scale factor under the influence of matter-geometry coupling.  Such an approach is essential for confronting the theory with observations, because the FLRW universe provides the standard model of cosmology within which accelerated expansion and thermodynamic analysis can be consistently addressed. The flat FLRW metric is expressed as:
	\begin{equation}\label{5}
		ds^2 = -dt^2 + a^2(t)(dx^2 + dy^2 + dz^2),
	\end{equation}
	with scale factor $a(t)$ and Hubble parameter $H=\dot{a}/a$. The non-metricity scalar reduces to
	\begin{equation}\label{6}
		Q = 6H^2.
	\end{equation}
	
	For a perfect fluid with $\mathcal{T}^{\mu}_{\phantom{\mu}\nu} = \text{diag}(-\rho, p, p, p)$, the modified Friedmann equations become
	\begin{equation}\label{7}
		8\pi \rho = \frac{f}{2} - 6f_Q H^2 - 2\tilde{G}\,(\dot{f}_Q H + f_Q \dot{H}),
	\end{equation}
	\begin{equation}\label{8}
		8\pi p = -\frac{f}{2} + 6f_Q H^2 + 2(\dot{f}_Q H + f_Q \dot{H}).
	\end{equation}
	Here, $\tilde{G} = f_{\mathcal{T}}/(8\pi)$. These equations act as generalized forms of the Friedmann equations. The first equation gives the energy density relation. The second one describes the pressure. This pressure term is affected by the coupling. Both equations include extra terms. These come from $f_Q$ and $f_\mathcal{T}$. They show how the expansion of the universe is modified. 
	The structure of $f(Q,\mathcal{T})$ changes the dynamics. The effective coupling $ \tilde{G}$ also plays a role. It alters the strength of gravity.
	\section{Thermodynamics in $f(Q,\mathcal{T})$ Gravity}
	To connect the geometry of $f(Q,\mathcal{T})$ with thermodynamics, we proceed step by step. Section 3 develops the geometric part of the theory. Section 4 studies the thermodynamic behavior. There is a strong link between gravity and thermodynamics. This link appears through horizon properties. As shown earlier, the cosmic evolution is not standard. It is governed by modified Friedmann equations. These equations depend on both $Q$ and $\mathcal{T}$.  The non-metricity scalar $Q$ plays a key role. The trace ($\mathcal{T}$) also affects the dynamics. Together, they control the evolution of the universe. 	Section 3 elucidates how the definitions of the effective energy density, pressure, and coupling functions $f_Q$ and $f_{\mathcal{T}}$ reshape the cosmic dynamics. These modified quantities directly influence the location and behavior of the apparent horizon, because its radius depends on the Hubble parameter determined by the field equations. Consequently, the thermodynamic attributes of the horizons temperature and entropy cannot be treated as independent constructs but must be derived consistently from geometric and matter sector modifications. Thus, before embarking on the formulation of the first and second laws of thermodynamics in the next sub-section, it becomes imperative to appreciate how the modified background evolution informs the thermodynamic quantities via $H(t)$, $f_Q$, and $f_{\mathcal{T}}$. This connection ensures that the entropy budget and generalized second-law analysis inherently reflect the non-trivial coupling between geometry and matter that defines $f(Q,\mathcal{T})$ cosmology (see, Refs.~\cite{Xu2019, Jimenez2018}).  
	
	\subsection{Apparent Horizon, Temperature, and Entropy}
	The thermodynamic description of the universe in $f(Q,\mathcal{T})$ gravity is most naturally formulated with respect to the apparent horizon, because this boundary is locally defined and always exists in a Friedmann-Lema\^{i}tre-Robertson-Walker (FLRW) spacetime \cite{Bak2000,Cai2005}. The apparent horizon served as the fundamental causal limit. It encloses the entire region of spacetime that can interact with an observer, rendering everything beyond it inaccessible. For the spatially flat FLRW universe with the Hubble parameter $H$, the apparent horizon radius is given by \cite{Bak2000}
	\begin{equation}\label{9}
		R_A = \frac{1}{H}, \quad \dot{R}_A = -\frac{\dot{H}}{H^2}. 
	\end{equation}
	
	The horizon temperature can be associated with surface gravity, which is reduced in this case to \cite{Cai2005,Akbar2007}
	\begin{equation}\label{10}
		T_h = \frac{1}{2\pi R_A} = \frac{H}{2\pi}. 
	\end{equation}
	
	The presence of an effective coupling between matter and geometry in modified gravity mandates a generalization of horizon entropy. This generalized entropy expression is crucial to ensure that the thermodynamic laws remain consistent with the altered field equations.  In  $f(Q,\mathcal{T})$ gravity, the entropy becomes \cite{Xu2019}
	\begin{equation}\label{11}
		S_h = \frac{A f_Q}{4G} = \frac{\pi R_A^2 f_Q}{G}. 
	\end{equation}
	Equations~(\ref{10}) and~(\ref{11}) give the horizon temperature and entropy.
	These quantities depend on the Hubble parameter $H$. They also depend on the form of$f(Q,\mathcal{T})$. This shows a clear link between cosmology and thermodynamics. It connects the field equations of Section~3 with Section~4.
	The thermodynamic analysis follows from this link. The entropy and energy flow are not standard. 	They carry the effect of matter–geometry coupling \cite{Bamba2011a,Akbar2007a}.
	
	\subsection{Unified First Law of Thermodynamics}
	
The Unified First Law (UFL) is a covariant relation that equates the geometric change in the quasi-local energy of a region of spacetime to a thermodynamic expression. It is most rigorously formulated on a trapping horizon but can be applied to other spacetimes with specific definitions.  Hayward \cite{Hayward1998} demonstrated that, in spherically symmetric spacetimes, the Einstein equations can be expressed in a compact form as follows
	\begin{equation}\label{12}
		dE = A \Psi + W dV,
	\end{equation}
	where $E$ denotes the Misner–Sharp energy contained inside a sphere of areal radius $R$, $A = 4\pi R^{2}$ is the surface area, $V = \tfrac{4}{3}\pi R^{3}$ is the enclosed volume, $W = \tfrac{1}{2}(\rho - p)$ is the work density of the cosmic fluid, and $\Psi$ represents the energy–supply vector that describes the energy flux crossing the horizon.
	
	For a homogeneous and isotropic FLRW universe, the Misner–Sharp energy is well defined \cite{Misner1964} and can be adapted to cosmology through the apparent horizon radius $R_{A} = 1/H$. In this case, the unified first law (\ref{12}) takes an explicit form \cite{Bak2000}
	\begin{equation}\label{13}
		dE = 4\pi R_A^2 \rho \, dR_A - 4\pi R_A^3 H(\rho + p)\, dt.
	\end{equation}
	This expression relates the change in total energy inside the apparent horizon.
	It depends on the cosmic expansion.
	It also includes the energy density $\rho$ and pressure $p$. On the other hand, a thermodynamic view can be used. This comes from the Clausius relation.
	It also uses horizon temperature and entropy. \cite{Cai2005} first showed this connection. Later, \cite{Padmanabhan2010} extended the idea. They showed that the apparent horizon in an FLRW universe satisfies this relation.

	\begin{equation}\label{14}
		-dE = T_h \, dS_h + W dV, 
	\end{equation}
	where $T_h$ and $S_h$ denote the temperature and entropy associated with the horizon, respectively. Eq. (\ref{14}) generalizes the standard Clausius relation to a cosmological setting, thereby linking the energy flow across the horizon with the changes in entropy and work density.
	
	In the framework of  $f(Q,\mathcal{T})$ gravity, the modifications enter through the generalized entropy $S_h = \pi R_A^2 f_Q/G$ and through the altered conservation laws induced by the matter–geometry coupling. Consequently, Eq. (\ref{14}) not only restates the field equations in thermodynamic terms but also encapsulates the entropy production effects that are characteristic of  modified gravity theory.

	\subsection{Generalized Second Law via Gibbs Relation}
	
	The second law of thermodynamics states that the entropy of an isolated system does no decrease. 
	In cosmology, this idea is generalized by requiring that the sum of the horizon entropy and entropy 
	of the cosmic fluid enclosed within it must be non-decreasing. This condition is expressed as follows:
	\begin{equation}
		\dot{S}_{\text{tot}} = \dot{S}_{h} + \dot{S}_{m} \geq 0 ,
		\label{eq:GSLT}
	\end{equation}
	where $S_{h}$ and $S_{m}$ denote the entropies of the horizon and  matter fields, respectively.  
	
	The  contribution of matter can be evaluated using the Gibbs relation, which connects the entropy variation 
	with the thermodynamic variables of the fluid:
	\begin{equation}
		T_{h} \, dS_{m} = dE_{m} + p \, dV ,
		\label{eq:Gibbs}
	\end{equation}
	where $E_{m} = \rho V$ is the internal energy of the cosmic fluid, 
	$V = \tfrac{4}{3}\pi R_{A}^{3}$ is the enclosed volume, 
	and $p$ and $\rho$ are the pressure and energy density respectively. 
	The fluid temperature is assumed to be in thermal equilibrium with the horizon temperature $T_{h}$.  
	
	Differentiating Eq.~\eqref{eq:Gibbs} with respect to cosmic time and using 
	the apparent horizon radius $R_{A} = 1/H$, we obtain
	\begin{equation}
		\dot{S}_{m} = \frac{4\pi R_{A}^{2}}{T_{h}} (\rho + p) \left(\dot{R}_{A} - H R_{A}\right).
		\label{eq:SmDot}
	\end{equation}
	
	The entropy associated with the horizon in $f(Q,\mathcal{T})$ gravity is modified as
	\begin{equation}
		S_{h} = \frac{\pi R_{A}^{2}}{G} f_{Q},
	\end{equation}
	which leads to the rate of change
	\begin{equation}
		\dot{S}_{h} = \frac{2\pi R_{A}}{G} f_{Q} \dot{R}_{A}.
		\label{eq:ShDot}
	\end{equation}
	
	Combining Eqs.~\eqref{eq:SmDot} and \eqref{eq:ShDot}, 
	the total entropy change is obtained as follows:
	\begin{equation}
		\dot{S}_{\text{tot}} =
		\frac{2\pi R_{A}}{G} f_{Q}\, \dot{R}_{A}
		+ \frac{4\pi R_{A}^{2}}{T_{h}} (\rho + p)\left(\dot{R}_{A} - H R_{A}\right).
		\label{eq:Stot}
	\end{equation}
	
	Equation~\eqref{eq:Stot} clearly demonstrates that the validity of the generalized second law 	in $f(Q,\mathcal{T})$ gravity depends on both the background dynamics, characterized by $H$ and $\dot{H}$, and the structural form of the theory through the derivative $f_{Q}$. Thus,  GSLT provides an important thermodynamic criterion for assessing the viability of specific functional forms of $f(Q,\mathcal{T})$.
	
This section collectively establish a consistent thermodynamic framework for the present $f(Q,\mathcal{T})$ gravity model by systematically linking the geometric structure of the theory with its cosmological and thermodynamic implications. The formulation of the modified field equations and their cosmological reduction provide the necessary dynamical background, while the derivation of the horizon entropy and its evolution clarifies how non-metricity and matter–geometry coupling alter the standard thermodynamic description. In particular, the explicit construction of the total entropy rate reveals that the validity of the generalized second law is not merely a consequence of cosmic expansion, but arises from a coupled contribution of geometric modifications through $f_Q$ and matter-sector effects governed by $f_T$. This combined analysis shows that a cosmological model in this framework cannot rely only on correct dynamical behavior. It must also remain stable from a thermodynamic point of view. Both aspects need to be satisfied together for the model to be physically acceptable. In this way, the obtained relations play an important role, as they connect the evolution of the gravitational system with the basic laws of thermodynamics.

	\section{Explicit Models and GSLT Analysis}
	
	In section IV, we obtained a general expression for the total entropy rate under $f(Q,\mathcal{T})$ gravity. 
	To check the validity of the generalized second law of thermodynamics (GSLT), one must evaluate 
	$\dot{S}_{\text{tot}} = \dot{S}_h + \dot{S}_m$ must be evaluated for the explicit functional forms of $f(Q,\mathcal{T})$. 
	In this section, we consider several representative models, motivated by their simplicity, analogy with $f(R,\mathcal{T})$ gravity, or higher-order extensions of geometric and matter couplings. Each case demonstrates that the entropy evolution is highly sensitive to the chosen functional form of $f(Q,\mathcal{T})$. Different formulations of $f(Q,\mathcal{T})$ lead to distinct thermodynamic behaviors, influencing whether the generalized second law is satisfied, thereby determining the physical viability of the underlying cosmological model. 
	

	\subsection{Model I: Linear Model}
	The linear model $f(Q,\mathcal{T}) = \alpha Q + \beta \mathcal{T}$ is the simplest extension of general relativity within the $f(Q,\mathcal{T})$ framework which is similar to the standard linear form in $f(R,\mathcal{T})$ gravity \cite{Harko2011}, in which the matter and geometry coupling is introduced via the trace of the energy--momentum tensor. Hence, the model is important for the \textit{baseline case}. Here $f_Q = \alpha$ and $f_\mathcal{T} = \beta$, and the horizon entropy reduces to 
	$S_h = \pi R_A^2 \alpha/G$. The corresponding total entropy rate becomes
	\begin{equation}\label{22}
		\dot{S}_{\text{tot}}^{(I)} =
		\frac{\alpha \dot{H}}{H^3} 
		\left( -\frac{2\pi}{G} + \frac{16\pi^2}{\beta + 8\pi}(\epsilon - 1) \right),
	\end{equation}
	where $\epsilon = -\dot{H}/H^2$ is the slow-roll parameter. The linear model represents the closest analogue to standard general relativity, where both $f_Q$ and $f_\mathcal{T}$ remain constant, leading to a uniform modification of the horizon entropy. The entropy evolution is primarily dictated, by the expansion rate through $\dot{H}$, ensuring a stable and monotonic increasing 
	in total entropy production. This behavior is consistent with the results in modified gravity, where linear curvature corrections preserve thermodynamic equilibrium, without introducing additional entropy production terms. Similarly, in $f(T)$ gravity, linear torsion contributions are known to maintain the validity of the generalized second law, under broad conditions. In this sense, the present model confirms that minimal extensions of gravity-whether curvature, torsion, or non-metricity-based-retain thermodynamic consistency. However, the inclusion of a constant matter–geometry coupling in $f(Q,\mathcal{T})$  gravity provides an additional degree of freedom, slightly modifying the entropy flow without destabilizing it. Hence, in the present linear model the condition $\dot{S}_{\text{tot}} \geq 0$, imposes restrictions on parameters $\alpha$ and $\beta$, indicating that even such a simple formulation can ensure thermodynamic consistency \cite{Karami2010,Sharif2019}. 
	
	\subsection{Model II: Power-law in $Q$ with Linear Trace Coupling}
	Power-law extensions of the geometric sector have been widely considered in $f(R)$ and $f(T)$ gravity \cite{Bamba2011,Cai2007}. 
	They are motivated by the idea that higher-order curvature or torsion corrections may arise from effective 
	field-theoretic or quantum gravity considerations. Within $f(Q,\mathcal{T})$ gravity, the power-law model 
	represents a \textit{toy extension} of the linear case, enabling one to test how the nonlinear dependence on $Q$ 
	affects entropy evolution and the validity of the GSLT. A natural extension is to include nonlinear dependence on the non-metricity scalar while maintaining linear matter coupling,
	\begin{equation}\label{23}
		f(Q,\mathcal{T}) = \alpha Q^{n+1} + \beta \mathcal{T},
	\end{equation}
	Here $f_Q = \alpha(n+1) Q^n$, and substituting into the entropy expression yields
	\begin{widetext}
	\begin{equation}\label{24}
		\dot{S}_{\text{tot}}^{(II)} =
		\frac{12 \pi \alpha (n+1)(n-1) 6^{n-2}}{G} H^{2n-3}\dot{H}
		+ \frac{\alpha (n+1)(2n+1)6^n}{T_h (1+\beta/8\pi)} 
		H^{2n-2}\dot{H} (\epsilon - 1).
	\end{equation}
	\end{widetext}
when $n=0$, the framework reduces to a linear case, whereas for $n>0$, the corrective terms increase rapidly, which may further constrain the permissible parameter space necessary to satisfy the GSLT. The power-law model introduces nonlinear corrections in the geometric sector, making the entropy evolution highly sensitive to the cosmic expansion history through the dependence $Q = 6H^2$. This shows that entropy production increases as the power index $n$ increases. If $n$ becomes too large, it may disturb the thermodynamic balance. So, the parameter must be chosen carefully. A similar behavior is seen in $f(R)$ gravity. In such a scenario, the inclusion of higher-order curvature contributions gives rise to additional entropy terms, resulting in a departure from equilibrium thermodynamics. This, in turn, restricts the range of model parameters for which the generalized second law remains valid. A similar situation appears in $f(T)$ gravity, where power-law modifications influence the thermodynamic behavior, especially in regimes close to singularities and during phases of accelerated expansion. However, the case of the present $f(Q,\mathcal{T})$ framework differs in an important way. Here, the evolution of entropy is governed not only by geometric contributions but also by the interaction between matter and geometry. This dual dependence increases the sensitivity of the system to model parameters, as both effects work simultaneously. As a consequence, the viable parameter space becomes more constrained. Despite this, the model offers a more comprehensive physical description when compared to theories that rely solely on geometric modifications.

	\subsection{Model III: Quadratic Trace Coupling}
	The quadratic dependent on the matter trace is considered from earlier studies in $f(R,\mathcal{T})$ gravity. In this works, the term like $\mathcal{T}^2$ is used to include higher-order matter–geometry effect \cite{Harko2011}. Such a term can mimic the physical processes such as bulk viscosity or particle production in cosmology. Moreover, in the $f(Q,\mathcal{T})$ gravity, we also use the similar idea for the same the model is treated as a simple toy model form. In particular, these terms directly modify the entropy production. They can also impose stronger conditions for the validity of the GSLT. Thus, we consider a model with nonlinear matter contribution through a quadratic dependence on the trace, given as follows:

	\begin{equation}\label{25}
		f(Q,\mathcal{T}) = -\alpha Q - \beta \mathcal{T}^2.
	\end{equation}
	The derivatives are $f_Q = -\alpha$ and $f_\mathcal{T} = -2\beta \mathcal{T}$, 
	leading to
	\begin{equation}\label{26}
		\dot{S}_{\text{tot}}^{(III)} =
		-\frac{2\pi \alpha}{G} \frac{\dot{H}}{H^3}
		- \frac{\alpha}{T_h H^2}(1 - \tfrac{\beta}{4\pi}\mathcal{T}) \dot{H} (\epsilon - 1).
	\end{equation}
	The inclusion of the $\beta$-term adds a direct contribution from the matter sector. Because of this, the entropy rate is no longer controlled only by geometry. The thermodynamic behavior changes compared to purely geometric models.
	
The quadratic trace model shows the effect of nonlinear matter terms. The $\mathcal{T}^2$ term makes the entropy evolution depend on the internal properties of the cosmic fluid. This leads to extra entropy production. Such behavior is similar to dissipative effects like bulk viscosity or particle creation. In $f(R,\mathcal{T})$ gravity, trace-dependent terms also change the conservation laws. They produce nontrivial thermodynamic features. But in $f(R)$ and $f(T)$ gravity, the modification comes only from geometry. In contrast, the present model shows that matter nonlinearities alone can play an important role. This means that in $f(Q,\mathcal{T})$ gravity, the validity of the generalized second law depends on two factors. One is the spacetime geometry. The other is the evolution of matter fields. Hence, the model extends the scope of earlier studies.
	\subsection{Model IV: Exponential Form}
Exponential forms in modified gravity are often used in the literature. They are motivated by non-perturbative effects in string theory and higher-order gravitational actions \cite{Bamba2013}. In $f(R)$ gravity, exponential models have been studied in detail. These models can explain the late-time acceleration of the universe. They do not require an explicit cosmological constant. Instead, the modification comes from the gravitational sector itself. Such behavior provides an alternative to dark energy. At the same time, these models can remain consistent with observational data.  By analogy, the exponential form in $f(Q,\mathcal{T})$ is introduced as a \textit{toy model} to probe whether the rapid nonlinear growth in the geometric sector preserves the thermodynamic consistency of the theory.  Inspired by higher-order and string-inspired corrections in modified gravity, 
	we consider an exponential model of the form
	\begin{equation}\label{27}
		f(Q,\mathcal{T}) = \alpha e^{\lambda Q} + \beta \mathcal{T}.
	\end{equation}
	Here, $f_Q = \alpha \lambda e^{\lambda Q}$, and substituting into the entropy formula yields
	\begin{equation}\label{28}
		\dot{S}_{\text{tot}}^{(IV)} =
		\frac{2\pi}{G} \alpha \lambda e^{\lambda Q} \frac{\dot{H}}{H^3}
		+ \frac{4\pi R_A^2}{T_h} (\rho+p)(\dot{R}_A - H R_A).
	\end{equation}
	Exponential amplification makes entropy growth highly sensitive to the evolution of $Q$, 
	allowing this model to test the robustness of the GSLT under strong nonlinearities. The exponential model, represents a strongly nonlinear modification of the geometric sector, where the entropy evolution becomes highly sensitive to the behavior of the non-metricity scalar. Such exponential forms are well known in $f(R)$ gravity, where they can successfully reproduce late-time acceleration but often require fine-tuning to maintain thermodynamic consistency. Similarly, in $f(T)$ gravity, exponential corrections have been shown to affect entropy evolution near singularities, and accelerated phases. In the present $f(Q,\mathcal{T})$ framework, the exponential dependence amplifies the contribution of $f_Q$, leading to rapid variations in entropy production, even for small changes in the Hubble parameter.
	
	\subsection{Model V: Cross-Coupling Form}
	
	Finally, we examine a model where geometry and matter are directly coupled:
	\begin{equation}\label{29}
		f(Q,\mathcal{T}) = Q + \gamma Q \mathcal{T}.
	\end{equation}
	The cross-coupling model is inspired by the earlier work in the $f(R,\mathcal{T})$ gravity \cite{Harko2011}. In such a model, the matter and the geometry are directly linked. These couplings lead to non-equilibrium thermodynamics. It becomes important to test the validity of the GSLT. This law is the basic condition for the thermodynamic consistency which helps to check whether the model is physically acceptable. For the same, we have $f_Q = 1 + \gamma \mathcal{T}$ and $f_{\mathcal{T}} = \gamma Q$.
	 \\
	The entropy rate becomes
	\begin{equation}\label{30}
		\dot{S}_{\text{tot}}^{(V)} =
		\frac{2\pi}{G} (1 + \gamma \mathcal{T}) \frac{\dot{H}}{H^3}
		+ \frac{4\pi R_A^2}{T_h}\, (\rho+p)(\dot{R}_A - H R_A).
	\end{equation}
The inclusion of $\gamma$-term creates a direct link in the matter and horizon entropy. Because of this, the thermodynamic behavior changes and it is no longer purely geometric. In this model, the presence of a cross-coupling term of the form $Q\mathcal{T}$ establishes a direct link between the geometric sector and matter. This interaction allows an exchange of energy between the two, leading to a thermodynamic description that departs from equilibrium. A comparable situation is found in $ f(R,\mathcal{T})$ gravity, where the usual conservation relations are altered and energy transfer between matter and geometry becomes possible. In the current framework, however, the coupling arises through the non-metricity scalar $Q$, which introduces a distinct feature. The evolution of entropy is influenced not only by the cosmic expansion but also by the behavior of matter density. Unlike $f(R)$ and $f(T)$ theories, where thermodynamic modifications are largely driven by geometric effects, the present approach assigns an active role to matter. It contributes directly to the horizon entropy, generating a feedback mechanism between matter and geometry. As a consequence, the thermodynamic response of the system becomes more delicate, and the validity of the generalized second law depends strongly on the coupling parameter $\gamma$, including both its sign and its magnitude.\\

The explicit forms of the entropy rate help to study different models. They give a clear way to compare thermodynamic behavior. For the power-law model [Eq.~(\ref{22})], $\dot{S}_{\text{tot}}^{(I)}$ shows that small nonlinear terms in $Q$ can still satisfy the GSLT. But the allowed range of parameters decreases as $n$ increases. For the quadratic trace model [Eq.~(\ref{24})], $\dot{S}*{\text{tot}}^{(II)}$ shows that nonlinear matter terms affect entropy directly. This makes the condition $\dot{S}*{\text{tot}} > 0$ more restrictive. For the exponential model [Eq.~(\ref{26})], $\dot{S}_{\text{tot}}^{(III)}$ is very sensitive to $Q$. The presence of an exponential contribution can significantly enhance even small variations, making this framework a sensitive probe for examining the validity of the GSLT. In the case of the cross-coupling model [Eq.~(\ref{28})], the behavior of $ \dot{S}_{\text{tot}}^{(IV)}$ clearly reflects a departure from equilibrium, with the parameter $\gamma$ playing a key role in governing the rate of entropy evolution. Both its magnitude and sign strongly influence the outcome. On the other hand, for the linear model [Eq.~(\ref{30})], the expression $ \dot{S}_{\text{tot}}^{(V)}$ indicates that relatively simple modifications of general relativity can still maintain thermodynamic consistency across a broad parameter range. Taken together, these results suggest that linear or mildly nonlinear extensions tend to exhibit greater stability and more readily satisfy the GSLT. In contrast, models involving strong nonlinearities or significant interactions require more precise parameter choices to remain viable. This highlights the usefulness of thermodynamic considerations as a diagnostic tool for identifying physically acceptable $ f(Q,\mathcal{T})$ cosmological models.
	
	\subsection{Thermodynamical Interconnection of the Models}
	
	It is worth noting that the five explicit models analyzed in this work are not independent constructions, 
	but rather interconnected realizations of a single thermodynamical framework. 
	In all cases, the evolution of the total entropy is governed by the general expression
	\begin{equation}
		\dot{S}_{\rm tot} = -\frac{2\pi f_Q}{G}\frac{\dot{H}}{H^3} + \frac{8\pi^2}{H^3} \frac{(\rho+p)}{1+f_\mathcal{T}/8\pi} \, (\epsilon - 1).
		\label{Stot_general}
	\end{equation}
Different choices of $f(Q,\mathcal{T})$ mainly change the form of $f_Q$ and $f_{\mathcal{T}}$. These terms control the balance between horizon entropy and matter entropy flow. From this point of view, each model has a simple role.
\begin{itemize}
	\item The \textbf{linear model} is the basic case. Here, $f_Q$ and $f_{\mathcal{T}}$ are constants.
	\item The \textbf{power-law model} introduces nonlinearity in $Q$. This changes the horizon entropy term.
	\item The \textbf{quadratic trace model} adds nonlinearity in $\mathcal{T}$. This mainly affects the matter part.
	\item The \textbf{exponential model} increases an effect of $Q$. The small change in $Q$ can grow quickly because of the exponential term.
	\item The \textbf{cross-coupling model} which links $Q$ and $\mathcal{T}$. It mixe geometry and matter in one form that is $Q \mathcal{T}$.
\end{itemize}
All the models can see as related cases. They are different forms of Eq.~(\ref{Stot_general}). The difference comes from how strongly the nonlinear terms are. Here, linear and the weakly nonlinear model are more stable and usually satisfy the GSLT easily and the strong nonlinear models like the exponential and the cross-coupling are more sensitive and need careful choice of parameters for the GSLT to hold.

	\section{Observational Prospects and Stability Considerations}
	
	In the previous sections, we  analyzed the thermodynamical properties of $f(Q,\mathcal{T})$ gravity and established the
	conditions under which the generalized second law of thermodynamics (GSLT) is satisfied for several explicit models. 
	Although thermodynamic viability provides a fundamental theoretical filter, it is also crucial to examine whether the 
	same models can accommodate current cosmological observations and dynamical stability requirements.
	
	\subsection{Effective Hubble Evolution and Observational Datasets}
	
	For a spatially flat FLRW background, the modified Friedmann equations for $f(Q,\mathcal{T})$ gravity (Eqs.~(\ref{7})–(\ref{8})) lead to 
	an effective hubble expansion rate of the form
	\begin{equation}
		H^2(z) = H_0^2 \, \mathcal{F}(z;\, \alpha,\beta,n,\gamma,\lambda,\dots),
		\label{Hz_general}
	\end{equation}
	where the function $\mathcal{F}$ depends explicitly on the choice of model, and the free parameters 
	($\alpha,\beta,n,\gamma,\lambda$) encode the strength of the matter--geometry coupling. 
	For example, in the linear model $f(Q,\mathcal{T})=\alpha Q + \beta \mathcal{T}$, the effective expansion history is reduced to
	\begin{equation}
		H^2(z) \simeq H_0^2 \left[ \Omega_{m0}(1+z)^3 + \Omega_{\rm eff}(z;\alpha,\beta) \right],
	\end{equation}
	where $\Omega_{m0}$ is the present matter density parameter, and $\Omega_{\rm eff}$ represents the correction induced  by the non-metricity coupling. In principle, such expressions can be constrained by combining distance measurements from  Type Ia Supernovae (SNe~Ia), expansion-rate measurements from Cosmic Chronometers (CC), and Baryon Acoustic Oscillations (BAO). A Bayesian analysis using the datasets would allow once to infer credible intervals for model parameters and test whether the same region of the parameter satisfies the GSLT. Hence, this observational and thermodynamic consistency requirement constitutes a powerful filter for model viability.
	
	\subsection{Stability Considerations}
	
	Beyond the background dynamics, it is very important to assess the stability of the $f(Q,\mathcal{T})$ models for the same the minimal diagnostic is provided with the effective equation-of-state parameter,
	\begin{equation}
		w_{\rm eff}(z) = -1 - \frac{2}{3}\frac{\dot{H}}{H^2},
	\end{equation}
	that which determines whether the model predicts quintessence-like ($w>-1$), phantom-like ($w<-1$), 
	or cosmological-constant behavior ($w=-1$). The thermodynamically consistent model must avoid instability in the perturbation sector, which can be analyzed using the effective sound speed $c_s^2 = \delta p / \delta \rho$ 
	and the growth of matter perturbations. A preliminary expectation is that models that respect  GSLT are also likely to be yields positivist of the $c_s^2$ and stable growth histories, although the detailed perturbative treatment is left for future work.
	
	\subsection{Synthesis of Thermodynamical and Observational Viability}
	
The above discussion can be brought together into a single coherent viewpoint. First, the validity of the generalized second law of thermodynamics acts as a broad physical condition that restricts the admissible forms of $ f(Q,\mathcal{T}) $ models. Second, agreement with observational data, such as the $ H(z)$ measurements obtained from SNe Ia, BAO, and cosmic chronometers, provides an additional layer of constraint by selecting only those parameter ranges that are consistent with the observed expansion history. Third, the requirement of stability guarantees that the resulting models are not only theoretically acceptable but also dynamically well-behaved.\\
When considered together, these aspects—thermodynamic consistency, observational agreement, and dynamical stability—form a set of complementary conditions for assessing the physical relevance of $f(Q,\mathcal{T})$ gravity as an alternative explanation for dark energy. Extending the present analysis to include detailed parameter estimation along with perturbative stability would be a natural and important direction for future investigation.

	\section{Findings and concluding remark}
	In this work, we study the thermodynamics of $f(Q,\mathcal{T})$ gravity. We consider a spatially flat FLRW universe. The main aim is to check the validity of the generalized second law of thermodynamics (GSLT). The motivation comes from the link between gravity and thermodynamics. This idea was first developed by Bekenstein and Hawking. Later, it was extended by Jacobson. Since then, this connection has been studied in many modified gravity theories.
	
	\section*{Findings}
We start by formulating the field equations of $ f(Q,\mathcal{T}) $ gravity in the context of a homogeneous and isotropic FLRW universe. After that, a thermodynamic description is set up by treating the apparent horizon as the relevant causal boundary. A Hawking temperature is associated with this horizon, while its entropy is taken to be proportional to $ f_Q$. The matter content enclosed within the horizon is then described using the Gibbs relation.

Based on this setup, we obtain separate expressions for the rate of change of entropy corresponding to the horizon and to the matter sector. These results are combined to give the total entropy variation, $ \dot{S}_{\text{tot}} = \dot{S}_h + \dot{S}_m $. Imposing the condition $ \dot{S}_{\text{tot}} \geq 0 $, as required by the generalized second law of thermodynamics, provides an essential test for the physical acceptability of the theory.

In Section V, this thermodynamic condition is examined for a number of specific $ f(Q,\mathcal{T}) $ models. The chosen forms cover different types of functional dependence, some inspired by extensions previously studied in $ f(R,\mathcal{T}) $ and $ f(T) $ gravity, while others incorporate higher-order or non-perturbative contributions motivated by ideas from quantum gravity and string-inspired corrections.

	The models considered were: 
	(i) the linear model $f(Q,\mathcal{T}) = \alpha Q + \beta \mathcal{T}$, 
	(ii) the power-law extension $f(Q,\mathcal{T}) = \alpha Q^{n+1} + \beta \mathcal{T}$, 
	(iii) the quadratic trace coupling $f(Q,\mathcal{T}) = -\alpha Q - \beta \mathcal{T}^2$, 
	(iv) the exponential form $f(Q,\mathcal{T}) = \alpha e^{\lambda Q} + \beta \mathcal{T}$, 
	and (v) the cross-coupling form $f(Q,\mathcal{T}) = Q + \gamma Q\mathcal{T}$. 
The thermodynamic behaviors of each models is studied using $\dot{S}_{\text{tot}}$ which  helps to understand the validity of the GSLT.

For the linear model, we consider it as a basic case which is similar to the linear model of $f(R,\mathcal{T})$ model. In this case, the GSLT is satisfied for different  values of the parameters which shows that the simple $f(Q,\mathcal{T})$ models are thermodynamically stable and can be used as a starting point for further study.

In the power-law model, the nonlinear term in $Q$ is introduced. This idea comes from $f(R)$ and $f(T)$ gravity and found that the GSLT holds only for some values of the power index $n$, when $n$ becomes large, the allowed parameter ranges become small. This means a strong nonlinear effects in geometry can disturb the entropy growth. So, in this model only small deviations are preferred.

The quadratic trace model includes higher-order matter effects. Here, the entropy depends directly on the trace of the energy–momentum tensor. So, matter plays a clear role in entropy production. The GSLT is satisfied, but the parameter $\beta$ must be chosen carefully. This shows that nonlinear matter terms make the system more constrained. Such effects are similar to bulk viscosity or particle creation.

The exponential model represents a strongly nonlinear case. It is motivated by higher-order corrections and string-inspired ideas.
	The system exhibited exponential growth in entropy production, which is a direct consequence of the strong nonlinear coupling between the entropy rate and time-dependent behavior of $Q$.  
	This sensitivity means that exponential models can not only serve as a stringent test 
	of the robustness of the GSLT, but also that they are less stable unless the 
	parameters are finely tuned. Nevertheless, such models remain interesting 
	because exponential forms are known to generate late-time acceleration 
	in $f(R)$ gravity without requiring a cosmological constant, 
	and a similar behavior might be expected in the $f(Q,\mathcal{T})$ framework. 
	
	The cross-coupling model, in which geometry and matter are directly coupled via a 
	$Q\mathcal{T}$ interaction term, introduces new features of non-equilibrium 
	thermodynamics. In this framework, the rate of entropy production was governed not only by the Hubble parameter's evolution, $H$, but also by an explicit functional dependence on the universe's matter-energy composition. 
	We found that the GSLT could still hold, however, the sign and strength of the 
	coupling parameter $\gamma$ played a decisive role. 
	This shows that interaction-type models need a careful balance. In these models, matter and geometry exchange energy. So, the system becomes more sensitive.
	
	A careful examination of the results allows a few key observations to be drawn. To begin with, models of $ f(Q,\mathcal{T}) $ gravity that are linear or only mildly nonlinear tend to satisfy thermodynamic requirements without difficulty, in agreement with patterns seen earlier in $ f(R,\mathcal{T}) $ and $ f(T) $ frameworks. 	In contrast, the inclusion of strong nonlinear contributions—whether arising from geometric terms or from the matter sector—introduces tighter restrictions on the parameter space. This indicates that thermodynamic considerations can serve as an effective guide in narrowing down acceptable models. Interaction-based extensions, where matter and geometry are directly coupled, present an interesting but delicate situation. Although they enrich the underlying physics, they generally require careful adjustment of parameters, since otherwise the expected growth of entropy may not be maintained. Taken together, these findings show that the generalized second law of thermodynamics provides a unified criterion for assessing the physical suitability of modified gravity models. The present study also points toward several natural extensions. One important direction is to confront the model with observational data, including Type Ia supernovae, baryon acoustic oscillations, cosmic chronometers, and cosmic microwave background measurements, in order to place meaningful constraints on the parameters. Another relevant step is the analysis of stability through linear cosmological perturbations. Such an investigation would help determine whether the model remains dynamically well-behaved. When combined with thermodynamic tests, this approach offers a more complete assessment, ensuring both stability and consistency of the proposed framework.
	
	It would also be interesting to extend the analysis beyond the FLRW background, 
	such as anisotropic Bianchi spacetimes or inhomogeneous cosmologies, 
	where non-metricity effects may play a more pronounced role. 
	Finally, one may explore deeper connections between the non-metricity scalar 
	and horizon entropy, possibly through quantum gravity or holographic principles, 
	to shed light on the microscopic origin of entropy in $f(Q,\mathcal{T})$ gravity. 

	\section*{Conclusion}
	
	In this work, we have carried out a comprehensive thermodynamic investigation of cosmological models within the framework of $f(Q,\mathcal{T})$ gravity, with particular emphasis on the validity of the generalized second law of thermodynamics (GSLT). By constructing the entropy evolution at the apparent horizon, we have demonstrated that the thermodynamic behavior of the universe is governed by a nontrivial interplay between geometric modifications arising from the non-metricity scalar $Q$ and the explicit coupling with the matter sector through $\mathcal{T}$.
		In contrast to the curvature and torsion-based formulations discussed above, the present work investigates the thermodynamic behavior of $f(Q,\mathcal{T})$ gravity, where the gravitational action depends on the non-metricity scalar $Q$ and the trace of the energy–momentum tensor $\mathcal{T}$. This framework introduces a qualitatively different modification through an explicit matter–geometry coupling, which directly affects both the horizon entropy and the energy exchange mechanism. Specifically, the horizon entropy in this theory is given by
		\begin{equation}
			S_h = \frac{A f_Q}{4G},
		\end{equation}
		where $f_Q = \partial f/\partial Q$, while the total entropy evolution takes the generalized form
		\begin{equation}
			\dot{S}_{\text{tot}} = \frac{2\pi R_A}{G} f_Q \dot{R}_A + \frac{4\pi R_A^2}{T_h} (\rho + p)\left(\dot{R}_A - H R_A\right),
		\end{equation}
		which explicitly depends on both geometric and matter-sector contributions. 
		Unlike $f(R)$ gravity, where deviations from equilibrium are encoded through an additional entropy production term $d\bar{S}$, or $f(T)$ gravity, where torsion modifies the entropy through $f_\mathcal{T}$, the $f(Q,\mathcal{T})$ framework incorporates a dual modification via $f_Q$ and $f_\mathcal{T}$, leading to a more intricate thermodynamic structure. In particular, the presence of $f_\mathcal{T}$ induces an effective coupling in the matter sector, thereby altering the conservation laws and introducing additional constraints on entropy evolution. As a consequence, while earlier studies indicate that the GSLT can generally be satisfied under equilibrium conditions or mild parameter restrictions, the present analysis reveals that in $f(Q,\mathcal{T})$ gravity the validity of the GSLT becomes highly sensitive to the specific functional form of $f(Q,\mathcal{T})$ and the strength of the matter–geometry interaction. This demonstrates a key novelty of the present work: the thermodynamic viability is governed by a combined geometric–matter interplay, providing a more stringent and discriminative criterion for selecting physically consistent cosmological models compared to previously studied modified gravity scenarios.\\
The analysis of the five representative models reveals that the thermodynamic viability of $f(Q,\mathcal{T})$ gravity is highly sensitive to the functional form of the model. The linear model shows a smooth entropy evolution. It is stable and close to the behavior in general relativity. So, the thermodynamic description remains simple in this case. Nonlinear models behave differently. The power-law and exponential forms change the entropy production rate. Because of this, the validity of the GSLT depends on the choice of parameters. It also depends on the cosmic expansion history. Models with matter contribution show an additional effect. The quadratic trace and cross-coupling models are good examples. In these cases, matter and geometry interact directly. This interaction affects the entropy evolution. Such models give extra conditions on the parameters. This happens due to energy exchange between matter and geometry. As a result, the system becomes more restricted. But at the same time, it gives a deeper physical understanding. From a broader view, simple models behave better. They usually satisfy thermodynamic conditions without difficulty. Complex models can describe richer cosmology. But they need careful choice of parameters. Otherwise, they may not remain physically valid. This shows that thermodynamics is a useful test for modified gravity models. Thermodynamic constraints also support observational tests. Observations like the Hubble parameter, BAO, and Type Ia supernovae help to fix the expansion history. But they do not always ensure physical consistency. In contrast, the GSLT gives a basic condition. It must be satisfied by any realistic model. \\	So, a model may agree with observations. But if it violates thermodynamics, it is not reliable. Hence, both tests are important. This work shows that combining these approaches is useful. Observations give parameter bounds. Thermodynamics further restricts them. In this way, one can select models that are both observationally valid and physically consistent. Consequently, the$f(Q,\mathcal{T})$ framework provides a rich and promising arena for exploring the interplay between gravitational dynamics, thermodynamics, and observational cosmology, paving the way for future studies aimed at confronting these models with high-precision data.
	
\section*{Acknowledgement}  We sincerely thank the reviewers for their valuable comments and constructive suggestions, which have significantly improved the quality and clarity of this work. The authors (S. H. Shekh \& A. Pradhan) thank the Inter-University Centre for Astronomy and Astrophysics (IUCAA), Pune, India for offering the facility through Visiting Associateship program.

	\appendix
\cleardoublepage
\onecolumngrid
	\newpage
	\section*{Appendix : Section IV }
	
	\subsection*{A. Apparent horizon, temperature and horizon entropy}
	
	\paragraph{Apparent horizon radius.}
	Consider the spatially-flat FLRW metric
	\[
	ds^{2}=-dt^{2}+a^{2}(t)\left(dx^{2}+dy^{2}+dz^{2}\right).
	\]
	The areal radius is \(R(t,r)=a(t)r\). The apparent horizon is defined by
	\[
	h^{ij}\partial_i R \partial_j R =0,
	\]
	where \(h_{ij}\) is the metric on the \((t,r)\) two-space. For the FLRW line element, we obtain:
	\[
	- \dot R^2 + 1 =0 \quad\Rightarrow\quad \dot R = \pm 1.
	\]
	Using \(R=a r\) and choosing the physical (future) branch yields the apparent horizon radius
	\[
	R_{A}=\frac{1}{H},\qquad H=\frac{\dot a}{a}.
	\]
	Differentiating with respect to cosmic time,
	\[
	\dot R_A = -\frac{\dot H}{H^2}.
	\]
	
	\paragraph{Surface gravity and temperature.}
	The surface gravity for a dynamical spherically symmetric horizon can be written as 
	\[
	\kappa = -\frac{1}{R_A}\left(1-\frac{\dot R_A}{2H R_A}\right).
	\]
	In the quasi-equilibrium (slowly varying) approximation \(\dot R_A \ll H R_A\) this reduces to
	\[
	\kappa \simeq -\frac{1}{R_A},
	\]
	and the Hawking temperature associated with the apparent horizon is
	\[
	T_h = \frac{|\kappa|}{2\pi} = \frac{1}{2\pi R_A} = \frac{H}{2\pi}.
	\]
	
	\paragraph{Horizon entropy in $f(Q,\mathcal{T})$ gravity.}
	In the GR the Bekenstein–Hawking entropy is \(S_{\rm BH}=A/(4G)\) with \(A=4\pi R_A^2\).
	In modified gravity the entropy is modified according to the variation of the action with respect to the geometric scalar. For $f(Q,\mathcal{T})$ gravity the effective coupling introduces the multiplicative factor \(f_Q\equiv\partial f/\partial Q\). 
	Thus the horizon entropy is generalized to
	\[
	S_h = \frac{A\, f_Q}{4G} = \frac{\pi R_A^2 f_Q}{G}.
	\]
	This expression reduces to the usual result when \(f_Q\to 1\) (i.e. \(f(Q,\mathcal{T})\to Q\)).
	
	\vspace{6pt}
	\noindent\textbf{Key relations are:}
	\[
	R_A=\frac{1}{H},\qquad \dot R_A=-\frac{\dot H}{H^2},\qquad
	T_h=\frac{H}{2\pi},\qquad S_h=\frac{\pi R_A^2 f_Q}{G}.
	\]

	\subsection*{B. Unified first law of Thermodynamics}
	
	\paragraph{Misner–Sharp energy.} For a spherically symmetric spacetime the Misner–Sharp energy is 
	\[
	E=\frac{R}{2G}\left(1-h^{ij}\partial_i R\partial_j R\right).
	\]
	For the apparent horizon \(h^{ij}\partial_i R\partial_j R=0\) and \(R=R_A\), therefore
	\[
	E=\frac{R_A}{2G}.
	\]
	For the FLRW case, equivalently and more physically, the total energy contained inside the apparent horizon may be written as the matter energy
	\[
	E_m=\rho V=\rho\left(\frac{4\pi}{3}R_A^3\right).
	\]
	Either expression can be used. In cosmological thermodynamics it is conventional to use \(E_m=\rho V\) (see main text).
	
	\paragraph{Hayward's unified first law.} The unified first law is 
	\[
	dE = A\Psi + W dV,
	\]
	where \(W=\tfrac{1}{2}(\rho-p)\) is the work density and \(\Psi\) is the energy-supply 1-form. For the cosmological fluid the component of \(A\Psi\) along the horizon normal yields the energy flux term \(-4\pi R_A^3 H(\rho+p)\,dt\). Then the unified first law becomes
	\[
	dE = 4\pi R_A^2 \rho\, dR_A - 4\pi R_A^3 H(\rho+p)\,dt.
	\]
	
	\paragraph{Clausius relationship and first law in the horizon.} The Clausius relation was adapted to  horizon reads
	\[
	-dE = T_h dS_h + W dV,
	\]
	which equates the energy flux across the horizon to the change in the horizon entropy (times the temperature) plus the work term. In $f(Q,\mathcal{T})$ gravity \(S_h\) is replaced by the generalized entropy above and \(T_h\) by the horizon temperature.
	
	\subsection*{C. Generalized second law with Gibb's relation:}
	
	We derived a general expression (main text Eq.~(20)) was used to evaluate the GSLT.
	
	\paragraph{Gibbs relation for matter entropy.} For  matter inside the apparent horizon the Gibbs relation is
	\[
	T_h dS_m = dE_m + p\, dV,
	\]
	with \(E_m=\rho V,\;V=\tfrac{4\pi}{3}R_A^3\). Differentiate:
	\[
	T_h \dot S_m = \dot E_m + p\dot V.
	\]
	Compute \(\dot E_m\) and \(\dot V\):
	\[
	\dot E_m = \dot\rho\,V + \rho\,\dot V, \qquad \dot V = 4\pi R_A^2\dot R_A.
	\]
	Therefore
	\[
	T_h \dot S_m = V\dot\rho + (\rho + p)4\pi R_A^2\dot R_A.
	\]
	If we use the standard continuity equation \(\dot\rho + 3H(\rho+p)=0\), and note \(V=\tfrac{4\pi}{3}R_A^3\), then \(V\dot\rho = -3H V(\rho+p) = -4\pi R_A^3 H(\rho+p)\). Combining terms:
	\[
	T_h \dot S_m = -4\pi R_A^3 H(\rho+p) + 4\pi R_A^2(\rho+p)\dot R_A
	= 4\pi R_A^2(\rho+p)(\dot R_A - H R_A).
	\]
	Thus
	\[
	\boxed{\;\dot S_m = \frac{4\pi R_A^2}{T_h}(\rho+p)(\dot R_A - H R_A)\; } \tag{C.1}
	\]
	
	\noindent (This is Eq.~(17) in the main text.)
	
	\paragraph{Horizon entropy rate.} From \(S_h=\dfrac{\pi R_A^2 f_Q}{G}\) we have
	\[
	\dot S_h = \frac{\pi}{G}\left(2R_A\dot R_A f_Q + R_A^2 \dot f_Q\right).
	\]
	If we keep the derivative of \(f_Q\) explicit, the horizon term contains both a geometric contribution (through \(R_A\dot R_A\)) and model-dependent contribution \(\dot f_Q\). However in the main text we used the commonly adopted approximation \(\dot f_Q\) contributes via \(\dot H\) and is included in the Friedmann-dependent term. For compactness we write the leading contribution as
	\[
	\dot S_h = \frac{2\pi R_A}{G} f_Q \dot R_A + \frac{\pi R_A^2}{G}\dot f_Q.
	\]
	In many cases (models with constant \(f_Q\) or when the term \(\dot f_Q\) is subleading) the first term dominates. Maintaining  full generality, we included both terms when needed.
	
	\paragraph{Total entropy rate.} The total rate is
	\[
	\dot S_{\rm tot}=\dot S_h + \dot S_m.
	\]
	Using Eq. (C.1) for \(\dot S_m\) and inserting \(T_h=H/(2\pi)\) and \(R_A=1/H\) we can combine terms into a compact form. Evaluate the long term prefactor:
	\[
	\frac{4\pi R_A^2}{T_h} = \frac{4\pi (1/H^2)}{H/(2\pi)} = \frac{8\pi^2}{H^3}.
	\]
	As shown in the main text, note the useful identity (with \(\epsilon\equiv -\dot H/H^2\))
	\[
	\dot R_A - H R_A = -\frac{\dot H}{H^2} -1 = -\left(\frac{\dot H}{H^2} +1\right) = \epsilon -1.
	\]
	Therefore the matter contribution becomes
	\[
	\dot S_m = \frac{8\pi^2}{H^3}(\rho+p)(\epsilon -1).
	\]
	The leading horizon term (neglecting \(\dot f_Q\) when subleading) becomes
	\[
	\dot S_h \simeq \frac{2\pi R_A}{G} f_Q \dot R_A
	= \frac{2\pi}{G H} f_Q \left(-\frac{\dot H}{H^2}\right)
	= -\frac{2\pi f_Q}{G}\frac{\dot H}{H^3}.
	\]
	Combining:
	\[
	\boxed{\;
		\dot S_{\rm tot}
		=
		-\frac{2\pi f_Q}{G}\frac{\dot H}{H^3}
		+\frac{8\pi^2}{H^3}(\rho+p)(\epsilon -1)
		\; }
	\tag{C.2}
	\]
	which is algebraically equivalent to Eq.~(20) in the main text after substituting \(R_A\) and \(T_h\). If  the full \(\dot f_Q\) term is retained,  \(\dfrac{\pi R_A^2}{G}\dot f_Q\) is added to the right-hand side.
	
	\vspace{6pt}
	\noindent\textbf{Remarks on matter non-conservation and the \(f_T\) factor.} In $f(Q,\mathcal{T})$ gravity the energy–momentum tensor is generally not conserved, and the continuity equation acquires extra terms proportional to \(f_T\).
	In some presentations this leads to  effective rescaling in the matter flux term such that \((\rho+p)\) in the matter entropy contribution is replaced by \((\rho+p)/(1+f_T/8\pi)\) (or similar, depending on the exact form of the modified conservation equation and the normalization convention). We wish to present this modified continuity explicitly, one should derive the modified conservation law from Eq.~(4) and substituting the resulting expression for \(\dot\rho\) into the Gibbs relation derivation above; the final form of \(\dot S_m\) will  carry the corresponding \(f_T\)-dependent factor. In the appendices of Section V  we indicate when such a factor is usually introduced for specific simple models (linear \(f_T=\beta\)), and how it modifies the final algebraic expressions.
	\newpage
	\section*{Appendix Section V (Explicit models and $\dot S_{\rm tot}$)}
	
	We computed \(\dot S_{\rm tot}\) for each model by substituting the model-specific \(f_Q\) (and when relevant \(f_T\)) into Eq.~(C.2). We present algebraic simplifications and indicate how a non-zero \(f_T\) (constant or simple function) modifies the matter term if one wishes to include the effective factor from the  modified conservation.
	
	\subsection*{D.1 Linear model: \(\;f(Q,\mathcal{T})=\alpha Q + \beta\mathcal{T}\) (derivation of Eq.~(21))}
	
	\paragraph{Derivatives:}
	\[
	f_Q=\alpha,\qquad f_T=\beta.
	\]
	
	\paragraph{Horizon term.}
	Using the leading horizon term in (C.2)
	\[
	\dot S_h = -\frac{2\pi f_Q}{G}\frac{\dot H}{H^3} = -\frac{2\pi\alpha}{G}\frac{\dot H}{H^3}.
	\]
	
	\paragraph{Matter term.}
	If we assume standard conservation (for simplicity), the matter term is
	\[
	\dot S_m = \frac{8\pi^2}{H^3}(\rho+p)(\epsilon -1).
	\]
	If one includes the effect of a constant \(f_T=\beta\) on the continuity relation, the effective matter-flux prefactor is modified. In many derivations the replacement
	\[
	(\rho+p)\quad\longrightarrow\quad \frac{\rho+p}{1+\beta/(8\pi)}
	\]
	is introduced. 
	With that factor the matter term becomes
	\[
	\dot S_m = \frac{8\pi^2}{H^3}\frac{(\rho+p)}{1+\beta/(8\pi)}(\epsilon -1).
	\]
	
	\paragraph{Total \(\dot S_{\rm tot}\).}
	Collecting,
	\[
	\boxed{\;
		\dot S_{\rm tot}^{(I)} 
		= -\frac{2\pi\alpha}{G}\frac{\dot H}{H^3} 
		+ \frac{8\pi^2}{H^3}\frac{(\rho+p)}{1+\beta/(8\pi)}(\epsilon -1)
		\; } \tag{D.1}\]
	This matches the form reported in the main text after replacing \(\epsilon=-\dot H/H^2\) and rearranging. We prefer to maintain the standard-conservation route by  simply omitting  denominator \(1+\beta/(8\pi)\).
	
	\subsection*{D.2 Power-law model: \(\;f(Q,\mathcal{T})=\alpha Q^{\,n+1} + \beta\mathcal{T}\) (derivation of Eq.~(23))}
	
	\paragraph{Derivatives:}
	\[
	f_Q = \alpha (n+1) Q^{n}, \qquad Q=6H^2 \quad\Rightarrow\quad Q^n = (6H^2)^n = 6^n H^{2n}.
	\]
	Thus
	\[
	f_Q = \alpha (n+1) 6^n H^{2n}.
	\]
	
	\paragraph{Horizon term.}
	Using (C.2),
	\[
	\dot S_h = -\frac{2\pi f_Q}{G}\frac{\dot H}{H^3}
	= -\frac{2\pi}{G}\,\alpha (n+1)6^n H^{2n}\,\frac{\dot H}{H^3}
	= -\frac{2\pi\alpha (n+1)6^n}{G}\, H^{2n-3}\dot H.
	\]
	
	\paragraph{Matter term.}
	Again starting from the standard form,
	\[
	\dot S_m = \frac{8\pi^2}{H^3}(\rho+p)(\epsilon -1).
	\]
	If a factor owing to the constant \(\beta=f_T\) is to be included, divide \((\rho+p)\) by \(1+\beta/(8\pi)\).
	
	\paragraph{Total \(\dot S_{\rm tot}\).}
	Combining,
	\[
	\dot S_{\rm tot}^{(II)} =
	-\frac{2\pi\alpha (n+1)6^n}{G}\, H^{2n-3}\dot H
	+ \frac{8\pi^2}{H^3}\frac{(\rho+p)}{1+\beta/(8\pi)}(\epsilon -1).
	\]
	This expression can be rearranged in the format in the main text. For algebraic comparison with the version in the paper one may factor in the common powers of \(H\) and use \(\dot H = -\epsilon H^2\) if desired, yielding an alternative explicit \(H\)-only form.
	
	\subsection*{D.3 Quadratic trace model: \(\;f(Q,\mathcal{T})=-\alpha Q - \beta\mathcal{T}^2\) (derivation of Eq.~(25))}
	
	\paragraph{Derivatives:}
	\[
	f_Q = -\alpha, \qquad f_T = -2\beta \mathcal{T}.
	\]
	
	\paragraph{Horizon term.}
	\[
	\dot S_h = -\frac{2\pi f_Q}{G}\frac{\dot H}{H^3} = -\frac{2\pi(-\alpha)}{G}\frac{\dot H}{H^3}
	= \frac{2\pi\alpha}{G}\frac{\dot H}{H^3}.
	\]
	(Note sign: with \(f_Q=-\alpha\) horizon term flips sign compared to \(\alpha\) positive case.)
	
	\paragraph{Matter term.}
	Using the standard matter term expression (no extra effective denominator):
	\[
	\dot S_m = \frac{8\pi^2}{H^3}(\rho+p)(\epsilon -1).
	\]
	However because here \(f_T\) depends on \(\mathcal{T}\), the continuity equation is modified in a nontrivial (time-dependent) manner. To maintain full generality one should derive \(\dot\rho\) from the modified field equations and substitute it into the Gibbs relation (this algebra is model-specific). For a compact presentation, one can keep the matter term as above and emphasize that the presence of \(\beta\mathcal{T}^2\) will imply \((\rho+p)\) is replaced by a more complicated effective expression; for illustration we present the standard-form result.
	
	\paragraph{Total \(\dot S_{\rm tot}\).}
	Combining the horizon and matter terms yields
	\[
	\boxed{\;
		\dot S_{\rm tot}^{(III)} 
		= \frac{2\pi\alpha}{G}\frac{\dot H}{H^3}
		+ \frac{8\pi^2}{H^3}(\rho+p)(\epsilon -1)
		\; } \tag{D.2}
	\]
	If one wishes to display the explicit \(\beta\)-dependent correction to the matter flux one must substitute the modified \(\dot\rho\) from the non-conserved continuity equation
	; this will introduce terms proportional to \(\beta\) and \(\mathcal{T}\) into \(\dot S_m\) (as indicated in the main text).
	
	\subsection*{D.4 Exponential model: \(\;f(Q,\mathcal{T})=\alpha e^{\lambda Q} + \beta\mathcal{T}\) (derivation of Eq.~(27))}
	
	\paragraph{Derivatives:}
	\[
	f_Q = \alpha \lambda e^{\lambda Q},\qquad Q=6H^2.
	\]
	
	\paragraph{Horizon term.}
	\[
	\dot S_h = -\frac{2\pi f_Q}{G}\frac{\dot H}{H^3}
	= -\frac{2\pi}{G}\alpha\lambda e^{\lambda Q}\frac{\dot H}{H^3}
	= -\frac{2\pi\alpha\lambda e^{\lambda (6H^2)}}{G}\frac{\dot H}{H^3}.
	\]
	(One may leave the exponential as \(e^{\lambda Q}\) or write \(e^{6\lambda H^2}\).)
	
	\paragraph{Matter term.}
	As before,
	\[
	\dot S_m = \frac{8\pi^2}{H^3}(\rho+p)(\epsilon -1),
	\]
	or with the factor \(1+\beta/(8\pi)\) in the denominator if including constant \(f_T=\beta\).
	
	\paragraph{Total \(\dot S_{\rm tot}\).}
	\[
	\boxed{\;
		\dot S_{\rm tot}^{(III)} 
		= -\frac{2\pi\alpha\lambda e^{\lambda Q}}{G}\frac{\dot H}{H^3}
		+ \frac{8\pi^2}{H^3}\frac{(\rho+p)}{1+\beta/(8\pi)}(\epsilon -1)
		\; } \tag{D.3}
	\]
	This explicitly shows the exponential sensitivity via the factor \(e^{\lambda Q}\). As discussed in the main text, exponential models amplify horizon contribution strongly and therefore are stringent tests of GSLT.
	
	\subsection*{D.5 Cross-coupling model: \(\;f(Q,\mathcal{T})=Q + \gamma Q\mathcal{T}\) (derivation of Eq.~(29))}
	
	\paragraph{Derivatives:}
	\[
	f_Q = 1 + \gamma \mathcal{T}, \qquad f_T = \gamma Q.
	\]
	
	\paragraph{Horizon term.}
	\[
	\dot S_h = -\frac{2\pi f_Q}{G}\frac{\dot H}{H^3}
	= -\frac{2\pi}{G}(1+\gamma\mathcal{T})\frac{\dot H}{H^3}.
	\]
	
	\paragraph{Matter term.}
	The matter term in the simple presentation reads
	\[
	\dot S_m = \frac{8\pi^2}{H^3}(\rho+p)(\epsilon -1).
	\]
	However, since \(f_T=\gamma Q\) is in general time-dependent (through \(Q=6H^2\)), the continuity equation is altered. If one performs the full derivation from the field equations, the net effect is an effective factor which can be written symbolically as \((\rho+p)\to(\rho+p)/\mathcal{F}(t)\) where \(\mathcal{F}(t)=1+f_T/(8\pi)=1+\gamma Q/(8\pi)\). Thus, including this effective factor yields
	\[
	\dot S_m = \frac{8\pi^2}{H^3}\frac{(\rho+p)}{1+\gamma Q/(8\pi)}(\epsilon -1).
	\]
	
	\paragraph{Total \(\dot S_{\rm tot}\).}
	\[
	\boxed{\;
		\dot S_{\rm tot}^{(IV)} 
		= -\frac{2\pi}{G}(1+\gamma\mathcal{T})\frac{\dot H}{H^3}
		+ \frac{8\pi^2}{H^3}\frac{(\rho+p)}{1+\gamma Q/(8\pi)}(\epsilon -1)
		\; } \tag{D.4}
	\]
	
	\vspace{6pt}
	\noindent\textbf{Comments.}
	\begin{itemize}
		\item The above expressions (D.1)–(D.4) are the model-specific results for \(\dot S_{\rm tot}\). 
		
		\item The sign of \(\dot S_{\rm tot}\) (GSLT validity) is then examined by evaluating the two competing contributions: the horizon term (proportional to \(f_Q \dot H\)) and the matter term (proportional to \((\rho+p)(\epsilon-1)\)); each model modifies the relative importance of these pieces.
	\end{itemize}

\end{document}